# On the First Passage Time and Leapover Properties of Lévy Motions

T. Koren[a], A.V. Chechkin[b], and J. Klafter[a]

[a]School of Chemistry, Tel Aviv University, Tel Aviv 69978 Israel

[b]Institute for Theoretical Physics NSC KIPT, Akademicheskaya st. 1, Kharkov 61108 Ukraine

**Abstract**

We investigate two coupled properties of Lévy stable random motions: The first passage times (FPTs) and the first passage leapovers (FPLs). While, in general, the FPT problem has been studied quite extensively, the FPL problem has hardly attracted any attention. Considering a particle that starts at the origin and performs random jumps with independent increments chosen from a Lévy stable probability law $\lambda_{\alpha,\beta}(x)$, the FPT measures how long it takes the particle to arrive at or cross a target. The FPL addresses a different question: Given that the first passage jump crosses the target, then how far does it get beyond the target? These two properties are investigated for three subclasses of Lévy stable motions: (i) symmetric Lévy motions characterized by Lévy index $\alpha$ ($0 < \alpha < 2$) and skewness parameter $\beta = 0$, (ii) one-sided Lévy motions with $0 < \alpha < 1$, $\beta = 1$, and (iii) two-sided skewed Lévy motions, the extreme case, $1 < \alpha < 2$, $\beta = -1$.

**I. Introduction**

Lévy motions constitute a fundamental family of random motions generated by stochastic processes with stationary and independent increments. Examples of the Lévy family include, among others, the Brownian, Cauchy and Lévy-Smirnov processes. Since their introduction [1] Lévy motions have been investigated extensively both theoretically and experimentally. In fact, Lévy type statistics [2] turned out to be ubiquitous and is observed in various areas including: physics (chaotic dynamics, turbulent flows [3,4]), biology (heartbeats [5], firing of neural networks [6]), seismology (recordings of seismic activity [7]), electrical engineering (signal processing [8-10]), and economics (financial time series [11-13]).

For many years Brownian motion has served as the dominant model of choice for random noise in continuous-time systems. Its remarkable statistical properties, on the one hand, and its amenability to mathematical analysis, on the other, have led Brownian motion to become *the* model of continuous-time random motion and noise. However, Brownian motion is just a single example of the Lévy family. Furthermore, it is a very special and misrepresenting member of this family.

Amongst the Lévy family, the Brownian member is the only motion with continuous sample-paths. All other motions have discontinuous trajectories, exhibiting jumps. Moreover, the Lévy family is characterized by selfsimilar motions. Brownian motion is the only selfsimilar Lévy motion possessing finite variance – all other selfsimilar Lévy motions have an infinite variance.

In the present work we study two coupled properties of Lévy motions. The first one is the first passage time (FPT): How long it takes a random walker starting at the origin and performing independent jumps distributed according to a Lévy stable probability law to cross or arrive at a fixed target point. The second property is the first passage leapover (FPL): How large is the walker's leap over the target; namely, given that the first passage jump crosses the target, then how far does it get beyond the target? The FPL refers to the first landing of those jumps (flights) that crossed the target for the first time. While the distribution of FPTs has been investigated quite extensively for some types of Lévy motions [14-20], the related problem of the FPLs has not attracted much attention. The main goal of the present work is to investigate the FPTs and FPLs for three subclasses of Lévy random motions: (i) symmetric Lévy motions (ii) one-sided Lévy motions, and (iii) two-sided skewed Lévy motions, the extreme case. We present numerical results, find the asymptotic behavior of the FPT and FPL probability density functions (PDFs) and compare the results with theoretical findings.

The paper is organized as follows. In Sec.II we remind the reader of the general expressions for Lévy stable distributions, the meaning of the parameters characterizing them, and explain the numerical algorithm. In Secs. III, IV and V we study symmetric, one-sided and skewed Lévy motions, respectively. We end with conclusions.

## II. Lévy motions

Let us assume that a one-dimensional random Lévy motion $x(t)$ starts at $x = 0$ at $t = 0$, and that there is a target located at $x = a > 0$. For the numerical studies of the FPT and FPL properties we use a simple discrete-time representation of the Lévy stable process,

$$x(n) = \sum_{j=1}^{n} \xi_j . \qquad (1)$$

The FPT $\tau_d$ and the FPL $l_d$ are defined as follows:

$$\tau_d = n, \quad l_d = x_n - d . \qquad (2)$$

Here $n$ is the number of steps to first cross or arrive at the target, and $x(n)$ is the distance of the first crossing over the target from the origin. See figure 1 for a schematic description of a leapover event.

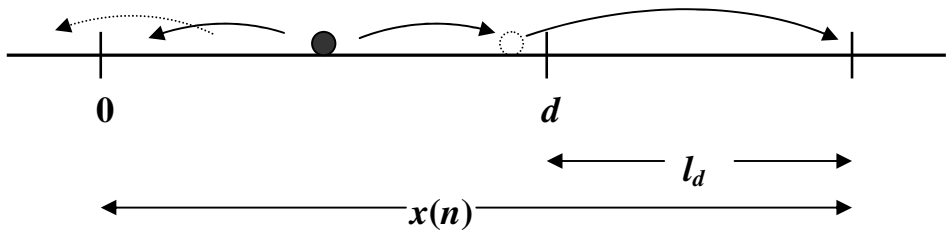

Figure 1. Schematic description of the leapover problem. A particle located initially at the origin $x=0$ performs a random walk according to a Lévy stable PDF. During the $n$-th step it crosses for the first time the target located at $x = d$. The $n$-th step corresponds to the crossing time $\tau_d$. The particle then lands at $x(n) = d + l_d$, where $l_d$ defines the leapover distance beyond the target.

Here the increments $\{\xi_j\}$ are chosen from a Lévy stable PDF $\lambda_{\alpha,\beta}(x;\mu,\sigma)$ expressed in terms of its characteristic function $\lambda_{\alpha,\beta}(k;\mu,\sigma)$ [21-23],

$$\lambda_{\alpha,\beta}(x;\mu,\sigma) = \int_{-\infty}^{\infty} \frac{dk}{2\pi} e^{-ikx} \lambda_{\alpha,\beta}(k;\mu,\sigma), \tag{3}$$

and

$$\lambda_{\alpha,\beta}(k;\mu,\sigma) = \exp\left[-\sigma^{\alpha} |k|^{\alpha}\left(1 - i\beta\frac{k}{|k|}\varpi(k,\alpha)\right) + i\mu k\right], \tag{4}$$

where

$$\varpi(k,\alpha) = \begin{cases} \tan\frac{\pi\alpha}{2}, & \text{if } \alpha \neq 1 \\ -\frac{2}{\pi}\ln|k|, & \text{if } \alpha = 1. \end{cases} \tag{5}$$

In general, the characteristic function and, respectively, the Lévy stable PDF are determined by the parameters: $\alpha, \beta, \mu$ and $\sigma$. The exponent $\alpha \in [0,2]$ is the Lévy index, $\beta \in [-1,1]$ is the skewness parameter, $\mu$ is the shift parameter, which is a real number, and $\sigma > 0$ is a scale parameter. The index $\alpha$ and the skewness parameter $\beta$ play a major role in our considerations, since the former defines the asymptotic decay of the PDF, whereas the latter defines the asymmetry of the distribution. These two properties of Lévy stable PDFs will be discussed in more detail below for each of the subclasses of Lévy stable random motions under consideration[1]. The shift and scale parameters play a lesser role in the sense that they can be eliminated by proper scale and shift transformations,

$$\lambda_{\alpha,\beta}(x;\mu,\sigma) = \frac{1}{\sigma}\lambda_{\alpha,\beta}(\frac{x-\mu}{\sigma};0,1). \tag{6}$$

Due to this property, in the present paper for brevity we denote the Lévy stable PDF $\lambda_{\alpha,\beta}(x;\mu,\sigma)$ by $\lambda_{\alpha,\beta}(x)$. We note the important symmetry property of the PDF, namely

---

[1] In the representation given by Eq.(4) the sign of $\beta$ is chosen so that the process with $\beta = 1$, $0 < \alpha < 1$ has *positive* increments, see Sec.IV below.

$$\lambda_{\alpha,-\beta}(x) = \lambda_{\alpha,\beta}(-x). \tag{7}$$

For instance, the asymptotics of the PDF $\lambda_{\alpha<1,1}(x)$, $x \to 0$ $(x>0)$ or $x \to \infty$ have the same behavior as $\lambda_{\alpha<1,-1}(x)$, $x \to 0$ $(x<0)$ or $x \to -\infty$. The property given by Eq. (7), obviously, leads to certain symmetry in formulating the FPT and FPL problems. We also note that only in three particular cases can the PDF $\lambda_{\alpha,\beta}(x)$ be expressed in terms of elementary functions: (i) $\alpha = 2$ (Gaussian distribution); in this case $\beta$ is irrelevant; (ii) $\alpha = 1$, $\beta = 0$ (Cauchy distribution); and (iii) $\alpha = 1/2$, $\beta = 1$ (Lévy-Smirnov distribution). In general, closed analytical forms of stable law PDFs are given via the Fox H-functions [23,24].

In what follows we investigate the FPT and FPL properties for three different Lévy motions, that is for three different statistics of the increments $\{\xi_j\}$: (i) symmetric Lévy stable PDFs, $\alpha \in (0,2], \beta = 0$; (ii) one-sided Lévy stable PDFs, $\alpha \in (0,1), \beta = 1$, and (iii) two-sided extremal Lévy stable PDF, $\alpha \in (1,2), \beta = -1$. The importance of Lévy motions with extremal values of skewness parameter $\beta$ is due to the fact that any Lévy stable process $x(t)$ can be written in the form $x_1(t) - x_2(t)$ where $x_1$ and $x_2$ are independent Lévy processes possessing the same Lévy index $\alpha$ and $\beta = \pm 1$ [25]. Examples of the three types of Lévy stable PDFs are shown in figure 2 for some particular values of $\alpha$.

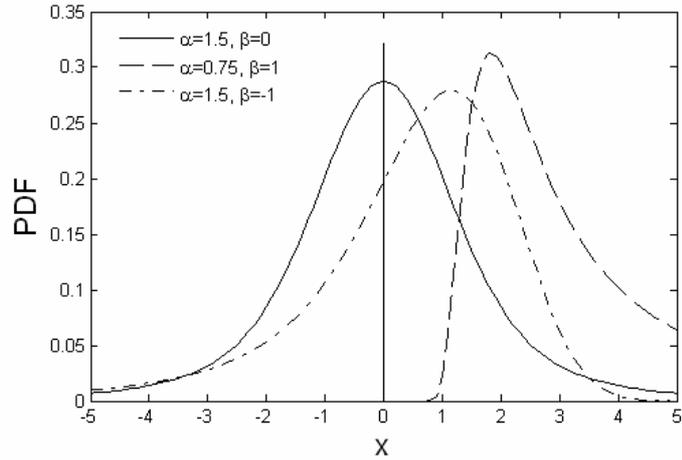

Figure 2. Lévy stable PDFs with different Lévy indices $\alpha$ and skewness parameters $\beta$. Examples of the three discussed subclasses are presented: symmetric ($\alpha = 1.5$, $\beta = 0$, solid line), one-sided ($\alpha = 0.75$, $\beta = 1$, dashed line) and two-sided skewed, the extreme case ($\alpha = 1.5$, $\beta = -1$, dashed-dotted line).

In the simulations the sets of increments $\{\xi_j\}$ are obtained by using the method of Chambers *et. al.* [21,22,26-28]. The process defined by Eq. (1) is repeated until $x(n)$ becomes larger or equal to the position of the target $x = d$ leading to two random values, $\tau_d$ and $l_d$, Eq. (2). The process in Eq. (1) then starts anew. In order to plot the asymptotic behavior of the FPT and FPL PDFs with a reasonable statistical accuracy, this procedure was repeated $1.5 \times 10^5$-$3.0 \times 10^5$ times for each random motion with fixed $\alpha$ and $\beta$.

**III. Symmetric Lévy motion**

We start by considering the particular case of Lévy motions derived from symmetric Lévy stable PDFs. The symmetric case corresponds to $0 < \alpha < 2$ and $\beta = 0$, with the characteristic function

$$\lambda_{\alpha,0}(k) = \exp(-|k|^\alpha) \quad . \tag{8}$$

The Cauchy PDF is a special case which corresponds to $\alpha = 1$,

$$\lambda_{1,0}(x) = 1/\pi\left(1 + x^2\right) \quad . \tag{9}$$

The symmetric Lévy stable PDF behaves asymptotically as [29,30]

$$\lambda_{\alpha,0}(x) \approx C_1(\alpha)/|x|^{1+\alpha} \quad , \quad x \to \pm\infty \quad , \tag{10}$$

where

$$C_1(\alpha) = \frac{1}{\pi}\sin(\pi\alpha/2)\Gamma(1+\alpha) \quad . \tag{11}$$

Typical trajectories of symmetric motions are shown in figure 3 for different Lévy indices $\alpha$.

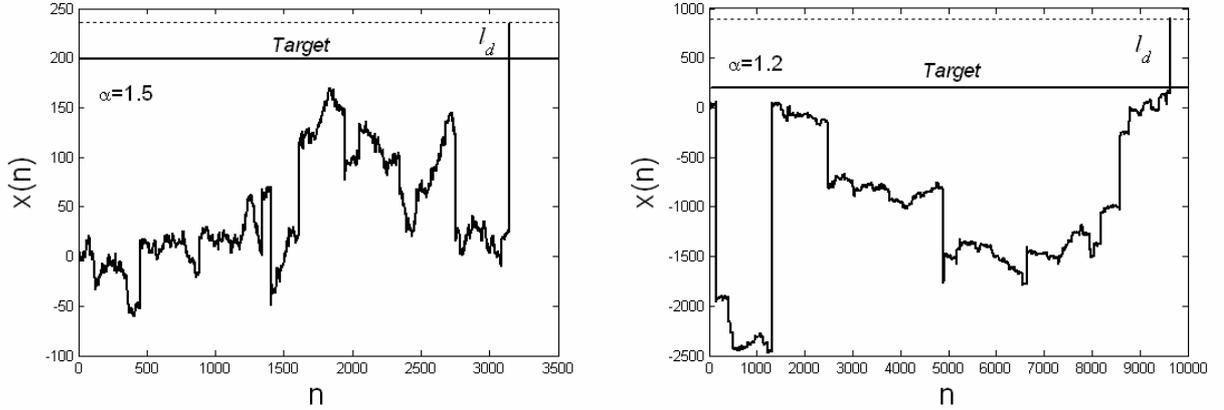

Figure 3. Trajectories obtained from numerical simulations with symmetric Lévy stable PDFs. As $\alpha$ becomes smaller larger jumps are more probable. In all cases the target is located at $d = 200$. Note the different scales for the different $\alpha$ values. The target location is shown by the full line and the leapover distance by the broken line.

It is quite obvious that the smaller $\alpha$ is the larger are the jumps and, as one might expect, the larger is also the size of leapover. To obtain the FPT law for symmetric Lévy motions we recall the fundamental result known as the "-3/2 law" or the Sparre-Andersen theorem [31,32]. The latter states that for any *discrete-time* random walk process with *independent* steps chosen from a continuous, *symmetric* arbitrary distribution, the FPT PDF decays asymptotically as $\sim n^{-3/2}$, where $n$ is the number of steps. Since in our problem the number of steps is equivalent to the current time, we obtain the asymptotics of the FPT PDF in the symmetric Lévy case,

$$p_\alpha(\tau_d) \propto \tau_d^{-3/2} \quad . \tag{12}$$

The FPT PDF decays with the same exponent independent of the Lévy index $\alpha$, $0 < \alpha \leq 2$ [14,16,33]. In particular for Brownian motion one obtains the first passage PDF

$$p_\alpha(\tau_d) = \frac{d}{\sqrt{4\pi D \tau_d^3}} \exp\left(-\frac{d^2}{4D\tau_d}\right) \quad , \tag{13}$$

which again decays as $\tau_d^{-3/2}$ for $\tau_d \gg d^2/(4D)$, where $D$ is the diffusion coefficient [14]. The "-3/2" exponent leads to the divergence of the mean first passage time (MFPT):

$$\langle \tau_d \rangle = \int_0^\infty \tau_d p_\alpha(\tau_d) d\tau_d = \infty \quad . \tag{14}$$

The "-3/2" behavior was obtained analytically and corroborated numerically for the special case in which an absorbing boundary is placed close to the location of the starting point of the Lévy motion [33]. Also note that in Ref. [33] the prefactor of the asymptotics given by Eq. (12) was obtained under the assumption of an exponential waiting time distribution of the single jumps. In figure 4 we show the FPT PDF on a log-log scale (natural logarithmic) for different values of the Lévy index $\alpha$ and the parameter $d$ (target location). The simulations nicely demonstrate the universal nature of Eq. (12).

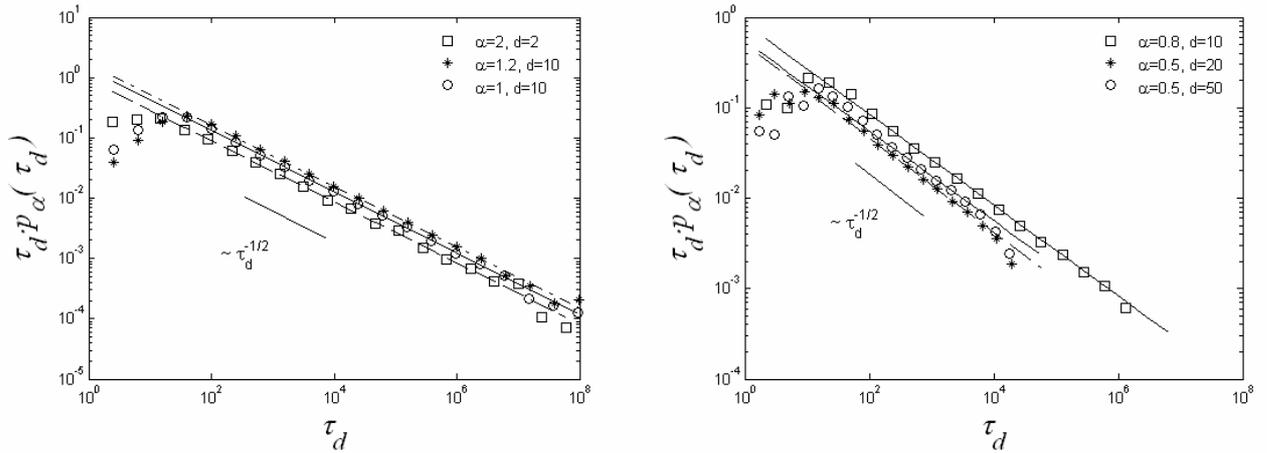

Figure 4. FPT PDFs $p_\alpha(\tau_d)$ of symmetric Lévy motions with different values of α. Plotted is $\tau_d \cdot p_\alpha(\tau_d)$ vs $\tau_d$ on a log-log scale. The solid lines correspond to the slope "-1/2" with the maximum relative error of 4%. The Sparre-Andersen behavior $p_\alpha(\tau_d) \sim \tau_d^{-3/2}$ is clearly observed.

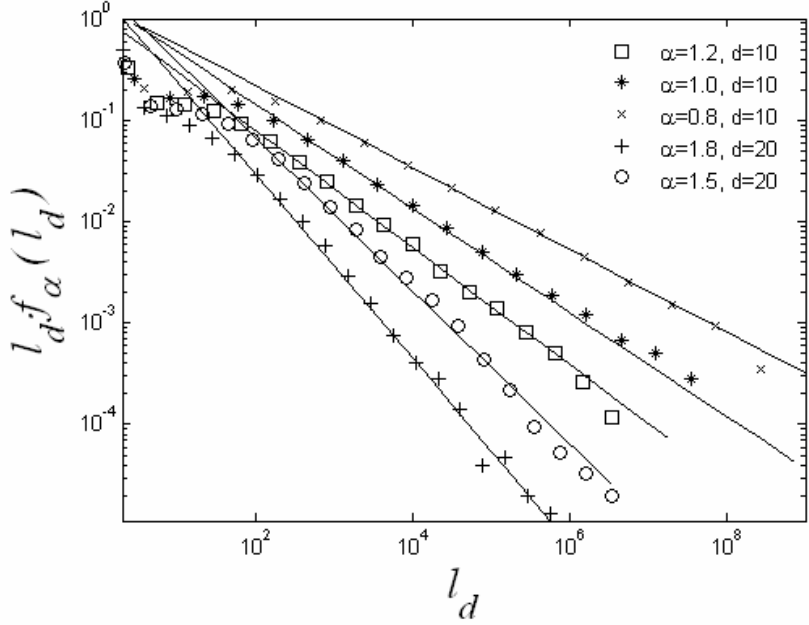

Figure 5. FPL PDFs $f_\alpha(l_d)$ of symmetric Lévy motions with different values of α. The target is located at different positions. Plotted is $l_d \cdot f_\alpha(l_d)$ vs $l_d$ on a log-log scale. The fits correspond to the asymptotic behavior of $f_\alpha(l_d) \sim l_d^{-(\alpha/2+1)}$ for the FPL PDF with the maximum relative error of 5%.

As mentioned, less is known about leapover properties. Of course for Brownian motion there is no leapover since the searcher approaches the target in a continuous way. The situation might be different for Lévy motions. For symmetric Lévy motions with index $\alpha$ our simulations support the following power-law decay of the FPL PDF in the limit of large $l_d$,

$$f_\alpha(l_d) \propto l_d^{-1-\alpha/2}, \quad 0 < \alpha < 2, \tag{15}$$

as shown in figure 5. For a phenomenological explanation of Eq. (15) we use the "superdiffusive" nature of a typical displacement $\delta x$ of the Lévy particle during time interval $\tau_d$,

$$\delta x \sim x \propto \tau_d^{1/\alpha}, \tag{16}$$

where $\tau_d$ has a PDF whose asymptotics is given by Eq. (12). Neglecting the constant shift in the leapover, which does not influence the far asymptotics, we write

$$l_d = x - d \propto \tau_d^{1/\alpha}. \tag{17}$$

Applying

$$f_\alpha(l_d) = p_\alpha(\tau_d)\frac{d\tau_d}{dl_d} \quad , \tag{18}$$

then, from Eqs. (17) and (12), by the simple change of variables, we arrive at the asymptotic behavior of the FPL PDF, Eq. (15). Surprisingly, the tail of the FPL PDF decays slower than the tail of the PDF of the increments, Eq. (10). However, our simple argument does not provide the prefactor which might be smaller than that of the tail of the increments PDF.

**IV. One-sided Lévy motion**

In this Section we proceed with the class of one-sided Lévy motions, which are also referred to as Lévy subordinators [17]. The increments of the one-sided motions are non-negative, and the Lévy stable PDF of the increments is non-zero on the positive semi-axis only. This class is characterized by the following Lévy indices and skewness parameter, respectively: $0 < \alpha < 1$, $\beta = 1$. Thus, the characteristic function, Eq. (4), reads as

$$\lambda_{\alpha,1}(k) = \exp\left[-|k|^\alpha \left(1 - i\frac{k}{|k|}\tan(\frac{\pi\alpha}{2})\right)\right] \quad , \tag{19}$$

or after transformation,

$$\lambda_{\alpha,1}(k) = \exp\left[-\frac{|k|^\alpha}{\cos(\pi\alpha/2)} e^{-i\,sign(k)\pi\alpha/2}\right] \quad . \tag{20}$$

The one-sided Lévy stable PDF is often characterized also via its Laplace transform [33],

$$\lambda_{\alpha,1}(s) = \int_0^\infty dx\, e^{-sx} \lambda_{\alpha,1}(x) \quad , \tag{21}$$

a representation which is equivalent to Eqs. (19) and (20),

$$\lambda_{\alpha,1}(s) = \exp\left[-\frac{s^\alpha}{\cos(\pi\alpha/2)}\right] \quad . \tag{22}$$

From Eq. (19) (or alternatively, Eqs. (20) or (22)) it follows that the increments PDF has the following asymptotic behavior ($x \to 0$) [35,36],

$$\lambda_{\alpha,1}(x) \approx C_2(\alpha) x^{-1-\mu(\alpha)/2} \exp\left[-C_3(\alpha) x^{-\mu(\alpha)}\right], \tag{23}$$

where

$$C_2(\alpha) = \frac{\alpha^{1/[2(1-\alpha)]} |\cos(\pi\alpha/2)|^{-1/[2(1-\alpha)]}}{\sqrt{2\pi |1-\alpha|}} \quad , \qquad (24)$$

$$C_3(\alpha) = |1-\alpha| \alpha^{\alpha/(1-\alpha)} |\cos(\pi\alpha/2)|^{-1/(1-\alpha)} \quad , \qquad (25)$$

and

$$\mu(\alpha) = \frac{\alpha}{|1-\alpha|} \quad . \qquad (26)$$

For $x \to \infty$ asymptotics of the increments PDF have the same behavior as that of the symmetric motion has, Eq. (10). The absolute values in Eqs. (24)-(26) are introduced for the sake of section V.

A particular case of the one-sided Lévy stable PDFs is the Lévy – Smirnov distribution, for which $\alpha = 0.5$, $\beta = 1$ [37],

$$\lambda_{1/2,1}(x) = \begin{cases} \dfrac{1}{\sqrt{2\pi}} x^{-3/2} \exp\left(-\dfrac{1}{2x}\right) & , \quad x > 0 \\ 0 & , \quad x < 0. \end{cases} \qquad (27)$$

This probability law appears, e.g., in the theory of crossing of a barrier by a Brownian particle, see Ref. [34], page 174. Trajectories of one-sided Lévy motions are shown in figure 6.

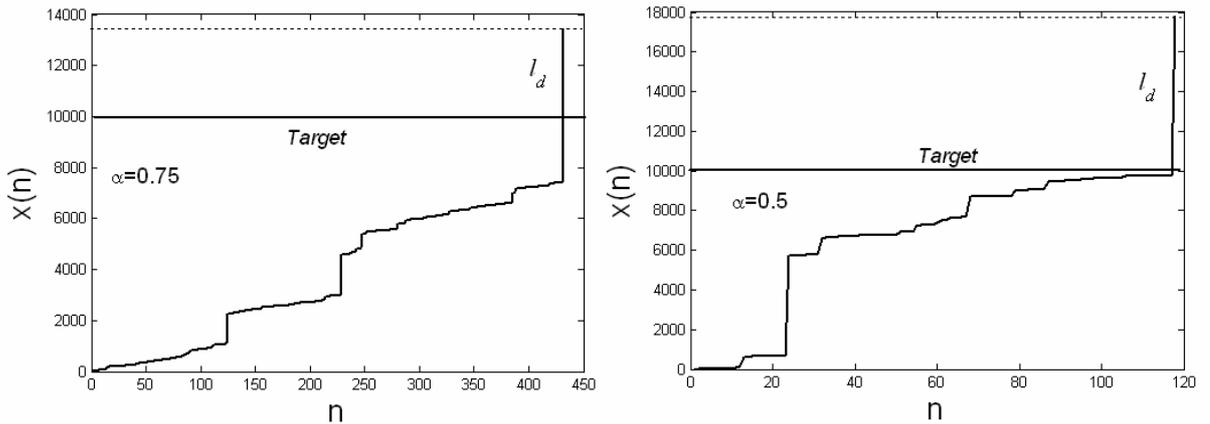

Figure 6. Trajectories of one-sided Lévy motions corresponding to different Lévy stable PDFs $(0 < \alpha < 1; \beta = 1)$. In both cases the target located at $d = 10^4$. The target location is shown by the full line and the leapover distance by the broken line.

The FPT and FPL problems for the one-sided Lévy motion were considered in Ref. [17]. We recall those results which are essential to our numerical analysis:

**1.** There exists a probabilistic relationship between the jumps $\xi_{\alpha,1}(x)$ of the one-sided Lévy motion and the FPTs $\tau_d$. No closed form could be derived for the FPT PDF for a general index α. The 'center case' in the one-sided Lévy motion, $\alpha = 0.5$, is the only case among the family of one-sided Lévy motions where an explicit expression for the FPT PDF has been derived. For further discussion we refer the readers to Ref. [17]. For the particular case $\alpha = 0.5$, the FPT PDF follows from Eq. (27),

$$p_{1/2}(\tau_d) = A\exp\left(-B\tau_d^2\right) \quad, \tag{28}$$

with $A = \sqrt{2/(d\pi)}$, $B = 1/(2d)$.

**2.** The MFPT is given for a general index $\alpha$, $0 < \alpha < 1$,

$$\langle \tau_d \rangle = \frac{d^\alpha}{\cos(\pi\alpha/2)\Gamma(1+\alpha)} \quad, \tag{29}$$

which scales with the target distance from the origin. Unlike the symmetric case in section III and the two sided skewed to be discussed in section V, here the MFPT is finite

**3.** The FPL PDF follows asymptotically the law of the increments,

$$f_\alpha(l_d) = \frac{\sin(\pi\alpha)}{\pi} \frac{d^\alpha}{l_d^\alpha(d+l_d)} \sim \frac{1}{l_d^{\alpha+1}}; \quad l_d \gg a \quad. \tag{30}$$

Comparisons between numerical simulations and the analytical results of the FPTs and FPLs are presented in figures 7-9, supporting the expressions in Eqs. (28-30).

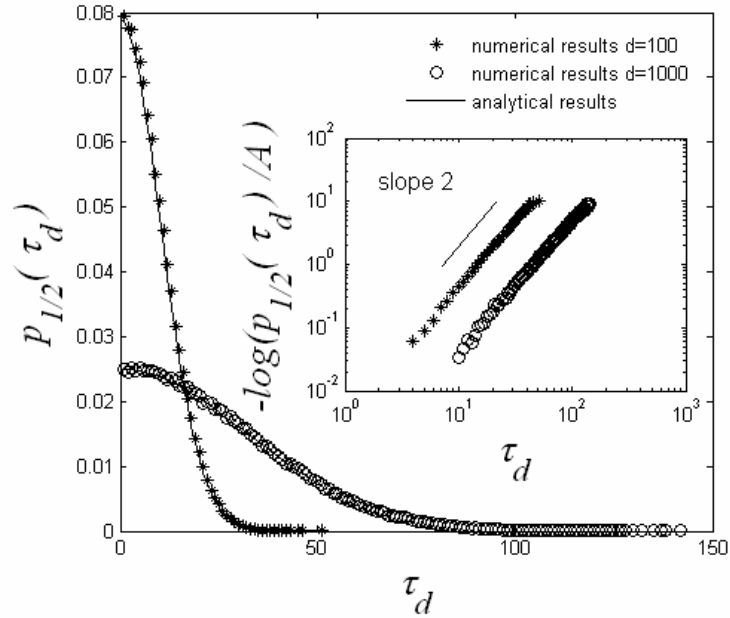

Figure 7. Comparison between analytical results corresponding to Eq. (28), and simulation results of the FPT for different target locations. Note that the inset displays $\log(-\log(p_{1/2}(\tau_d)/A))$ versus $\log \tau_d$ supporting Eq. (28).

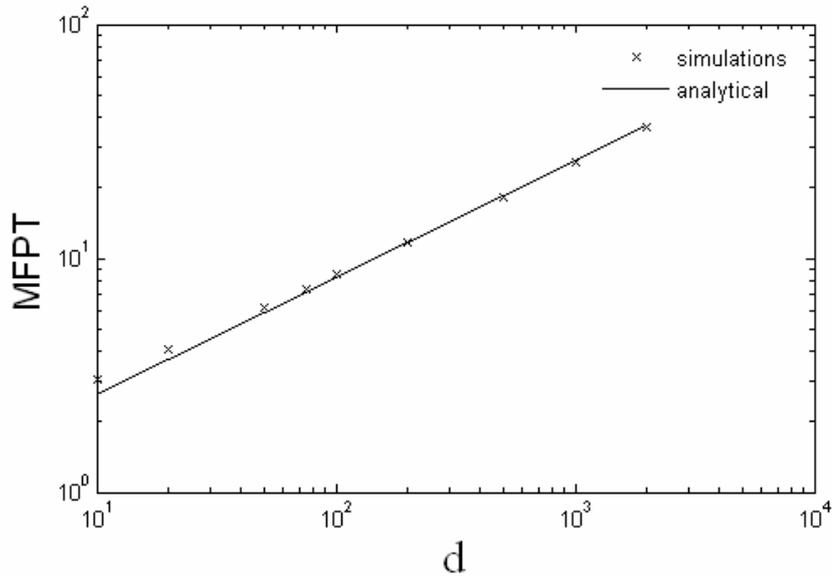

Figure 8. The MFPT for $\alpha = 0.5$ as a function of the target location showing a good agreement with Eq. (29) (the solid line). The relative error in the slope is 5%.

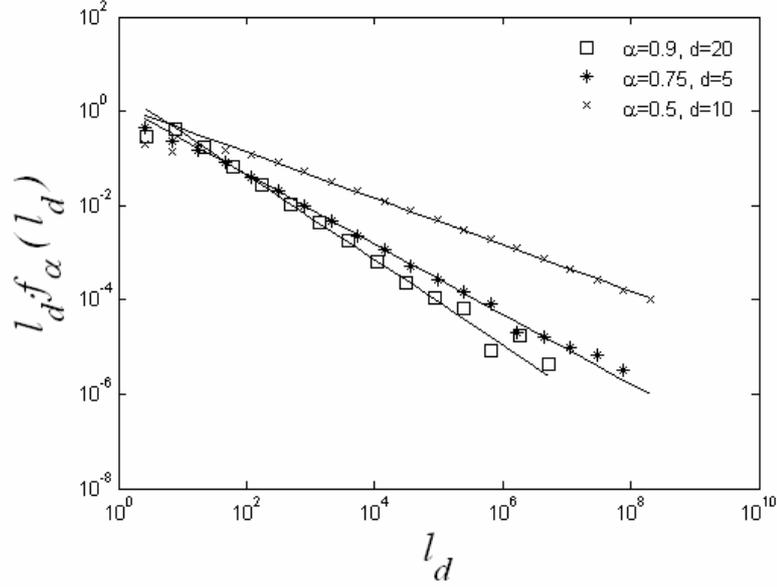

Figure 9. FPL PDFs $f_\alpha(l_d)$ of one-sided Lévy motions with different values of $\alpha$ and different target positions. Plotted is $l_d \cdot f_\alpha(l_d)$ vs $l_d$ on a log-log scale. The solid lines correspond to the fits supporting asymptotic behavior of $f_\alpha(l_d) \sim l_d^{-(\alpha+1)}$ with the maximum relative error of 1.8%.

## V. Two-sided skewed Lévy motion, the extreme case

Another interesting subclass of the Lévy motions is the extreme case of the two-sided but skewed Lévy motion of index $\alpha$, $1 < \alpha < 2$, and skewness parameter $\beta = -1$. This is a two-sided motion which possesses large jumps in the negative direction and small ones in the positive direction (see figure 2); thus, crossing continuously the target point located on the positive semi-axis, $d > 0$ (recall that at $t = 0$ the process begins at $x = 0$). Mathematically, continuous crossing implies that $x(\tau_d) = d$ with probability 1. However, in simulations, due to time discretization small leapovers occur which are negligible when compared with the large leapover values for $\alpha$, $1 < \alpha < 2$ in figure 5. Therefore, the leapover $l_d$ is practically zero, as in the Brownian case. The PDF of the increments in this case has power-law asymptotics on the negative semi-axis and falls off in an exponential way on the positive semi-axis [35,36],

$$\lambda_{\alpha,-1}(x) \approx \begin{cases} C_1(\alpha)/|x|^{1+\alpha}, & x \to -\infty \\ C_2(\alpha) x^{-1-\mu(\alpha)/2} \exp\left[-C_3(\alpha) x^{-\mu(\alpha)}\right], & x \to +\infty \end{cases}. \qquad (31)$$

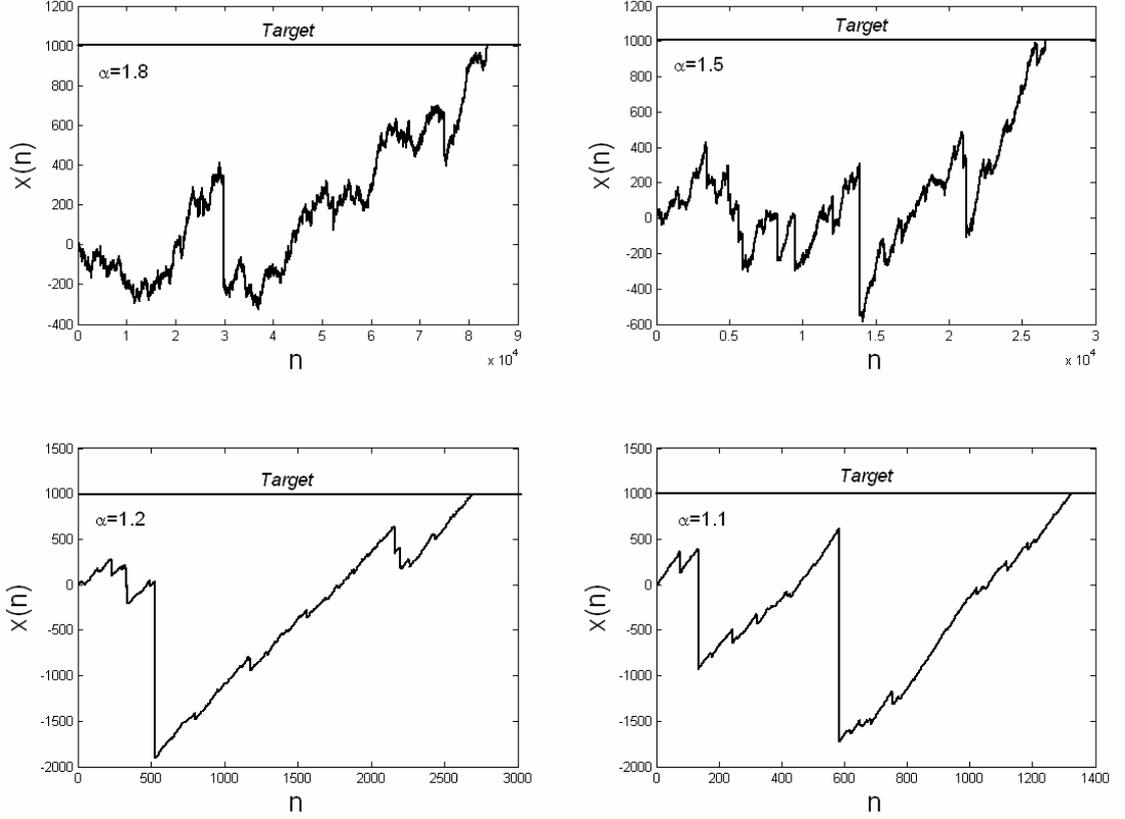

Figure 10. Trajectories obtained for different two-sided skewed distributions $(1 < \alpha < 2; \beta = -1)$. In this case the leapover is practically 0. The target located at $d = 10^3$.

where $C_1$, $C_2$ and $C_3$ and $\mu(\alpha)$ are given by Eqs. (11), (24), (25) and (26), respectively.

Typical trajectories are shown in figure 10 for different $\alpha$ values displaying large jumps away from the target and small ones towards it.

As to the FPT problem for skewed Lévy motions of index $\alpha$, a theorem was proven by Skorokhod [25] according to which the FPT $\tau_d$ is given by a one-sided Lévy stable PDF of the Lévy index $\alpha' = 1/\alpha$, $1/2 < \alpha' < 1$, the skewness parameter $\beta' = 1$, and the scale parameter

$$\sigma^{\alpha'} = a\left[\cos\left(\frac{\pi\alpha'}{2}\right)\right]\left[-\cos\left(\frac{\pi\alpha}{2}\right)\right]^{1/\alpha}. \qquad (32)$$

Namely, the FPT PDF is given by:

$$p_\alpha(\tau_d) = \frac{1}{2\pi} \int_{-\infty}^{\infty} dk \exp(-ik\tau_d) p_\alpha(k) , \qquad (33)$$

where,

$$p_\alpha(k) \equiv \lambda_{\alpha',1}(k;0,\sigma) = \exp\left\{-\sigma^{\alpha'}|k|^{\alpha'}\left(1-i\,sign(k)\tan\frac{\pi\alpha'}{2}\right)\right\} , \qquad (34)$$

so that asymptotically,

$$p_\alpha(\tau_d) \sim \frac{1}{\tau_d^{1/\alpha+1}} . \qquad (35)$$

The FPT PDF decays as a power law with the exponent $-1/\alpha - 1$ (compare with Eq. (12) and Eq. (28)). Since $1 < \alpha < 2$ the MFPT which corresponds to $p_\alpha(\tau_d)$ in Eq. (35) diverges.

The FPT PDF can not be written in a closed form, and numerical integration of Eqs. (33) and (34) is needed. We use the Quasi Monte Carlo (QMC) method which converged faster than the regular Monte Carlo in our case [38,39]. The results of the numerical simulations and numerical integrations for the FPT PDFs are presented in figures 11-13.

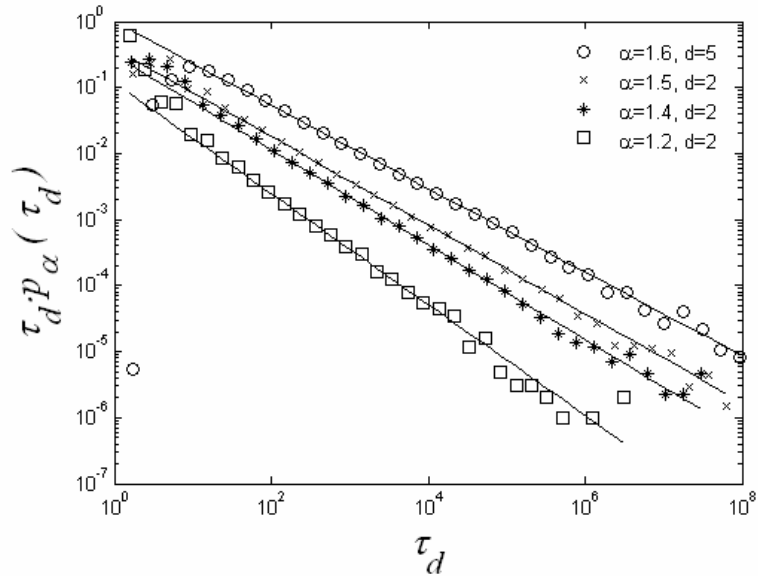

Figure 11. Plotted on a log-log scale is $\tau_d \cdot p_\alpha(\tau_d)$ vs $\tau_d$ of Lévy motions with $1 < \alpha < 2$, $\beta = -1$. The solid lines correspond to the behavior $\tau_d \cdot p_\alpha(\tau_d) \sim \tau_d^{-1/\alpha}$ which confirms the decay in Eq. (35). The maximum relative error is 1.25%.

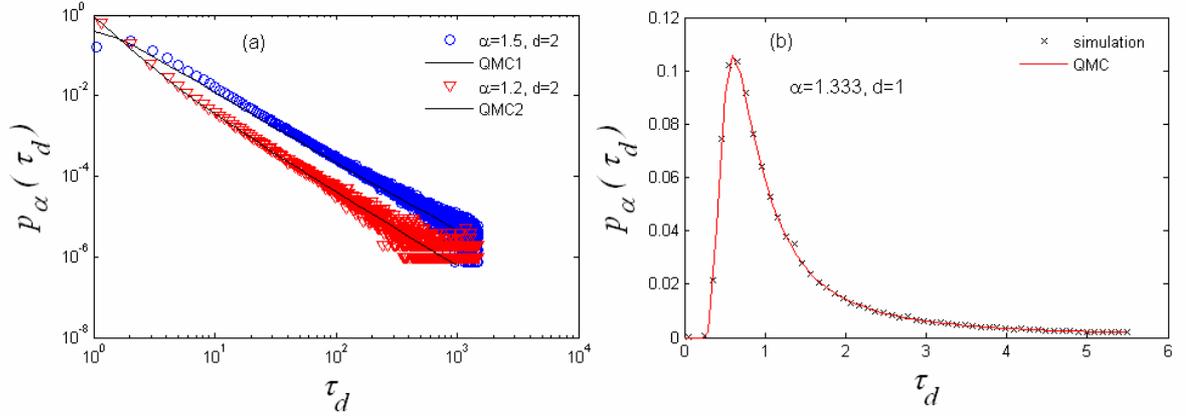

Figure 12. FPT PDF $p_\alpha(\tau_d)$ of Lévy motions with $1<\alpha<2$, $\beta=-1$ obtained from simulations and compared with numerical integrations (QMC) at long (a) and short (b) times, respectively. The agreement between the numerical integration of equations (33) and (34) and the simulations is clearly seen in whole domain of FPT.

Interestingly, the FPT PDFs of the symmetric Lévy motions decay asymptotically slower than the skewed case ($1<\alpha<2$, $\beta=-1$) discussed here. Therefore, the latter distributions, characterized by long jumps away from the target, lead to the seemingly paradoxical result of being a better strategy for reaching the target.

Finally, we speculate on the generalization of our FPL PDF of Lévy motions with $0<\alpha<1$ by allowing the skewness parameter $\beta$ to vary ($0<\beta<1$). The results presented in figure 13.

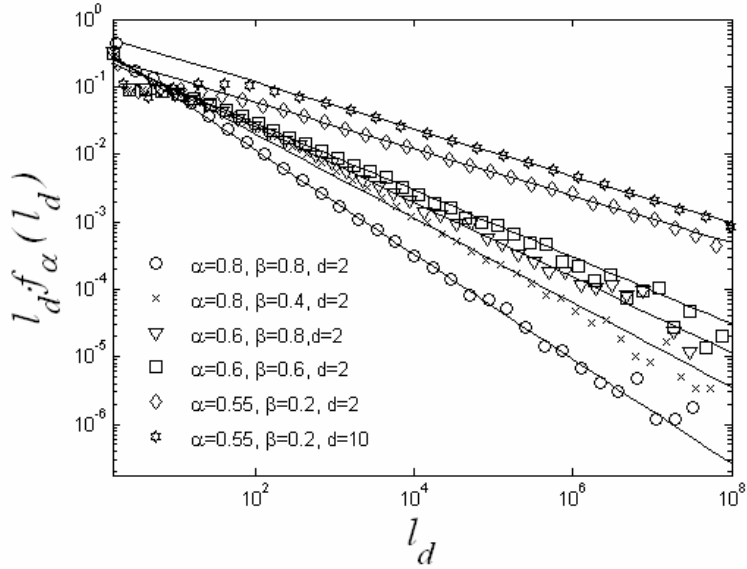

Figure 13. A log-log plot of FPL PDFs $f_\alpha(l_d)$ of Lévy motions with $0 < \alpha < 1$ and varying skewness $0 < \beta < 1$. The solid lines fit to $f_a(l_d) \sim l_d^{-(\alpha/2(\beta+1)+1)}$ with the maximum relative error of 6%.

## VI. Conclusions

In summary, we have presented calculation results of the first passage time and leapover PDFs for three subclasses of the Lévy motions, characterized by the different values of the Lévy index $\alpha$ and the skewness parameter $\beta$:

1. For the symmetric motion, $0 < \alpha < 2$, $\beta=0$, we demonstrated the universal decay of the FPT PDF with the exponent "-3/2". We found that the FPL PDF has a power-law asymptotics with the exponent $-1-\alpha/2$; thus, the decay is slower than that of the PDF of the increments of the Lévy motions.

2. For the one-sided Lévy motions with positive increments, $0 < \alpha < 1$, $\beta=1$, we provided detailed numerical simulation of the FPT problem for the particular case $\alpha = 0.5$, and confirmed numerically the power-law decay of the FPL PDF with the exponent $-1-\alpha$ (that is, the same decay as for the PDF of the Lévy increments) for different Lévy indices $\alpha$ and different target positions.

3. For the extreme case of the two-sided skewed Lévy motions $1 < \alpha < 2$, $\beta = -1$, we demonstrated that the FPT PDF is actually the one-sided Lévy stable PDF of Lévy index $\alpha' = 1/\alpha$ and skewness parameter $\beta' = 1$, thus the PDF decays asymptotically as a power law with the exponent $-1 - 1/\alpha$ larger than the exponent of "-3/2" obtained in the symmetric case.

4. Finally, based on numerical simulations of the general case of Lévy index $\alpha$, $0 < \alpha < 1$, and skewness parameter $\beta$, $0 \leq \beta \leq 1$, we suggest a power-law decay of the FPL PDF with the exponent $-1 - \alpha(\beta + 1)/2$, which reduces to the results for the symmetric, $\beta = 0$, and the one-sided, $\beta = 1$, cases, respectively.

**Acknowledgements**

The authors acknowledge discussions with Iddo Eliazar, Michael Andersen Lomholt and Ralf Metzler. AVC acknowledges hospitality of the School of Chemistry, Tel Aviv University.